# A Generic Web Component for WebRTC Pub-Sub

Kundan Singh
Intencity Cloud Technologies
San Francisco, CA, USA
kundan10@gmail.com

## ABSTRACT

We present video-io, a generic web component to publish or subscribe to a media stream in WebRTC (web real-time communication) applications. Unlike a call or conference room abstraction of existing video conferencing services, it uses a named stream abstraction, which is useful in many scenarios beyond just a call or conference. It keeps most of the application logic in the endpoint using the extensive application interface of this component, and keeps any vendor specific access control or signaling negotiation in a service-specific connector implementation. This allows an app developer to write once, and be able to run the web app on different servers or services. We also demonstrate its flexibility by implementing the connector for ten different existing systems and services. Decoupling the app from the hosted vendor service promotes innovation in the endpoint beyond what a single vendor locked client app can offer.

## 1. INTRODUCTION

Historically, communication systems had a clear separation between the endpoints and the network services. For phone, email, or even VoIP (voice-over-IP), you can get a client device or app, and connect to any service provider [1][2]. For web-based communication systems, the endpoint is not just the client browser but also the web app running in that browser. Unlike installed clients or apps, the emergence of SaaS (software-as-a-service) often promotes vendor lock-in [3][4] where the communication services can only be accessed by that vendor's own web apps. This hinders independent innovation in the client apps or endpoints.

Although WebRTC (web real-time communication) [5][6] (see Fig.1) enables audio/video flows between independent and competing browsers, it is often used by web developers and vendors to create islands of communication apps [7], where users of one web app cannot easily communicate with those of another. Unlike standards-based VoIP/SIP (session initiation protocol) [1][8] systems, a WebRTC [5] system has its own, often proprietary, signaling and control of the media flows, because they are intentionally left out of the specification [9][10]. A costly side effect is that the web apps are often tied to a specific vendor service, and web app developers end up creating separate apps for different services. Even for popular media servers or conferencing services, the APIs are so diverse that it is very hard to create a single web app connecting to multiple diverse services.

To solve this, as shown in Fig.1(c), we identify and separate the application logic that runs in the endpoint, from what depends on the vendor service, e.g., signaling negotiation, or user access control. We encapsulate the client-side implementation in a generic and reusable web component, named video-io. It can publish or subscribe to a named stream. This named stream abstraction can depend on and is implemented using a vendor service, but hides the vendor specific signaling from the client app, using a well defined API. Thus, a web app written once, can run on or connect to a wide range of vendor services.

Our video-io component is part of the RTC Bricks project [11], which includes several HTML5 web components to build a range of multimedia collaboration applications. It does not use any client-side JavaScript framework, and can integrate easily with any web app, installed progressive web app (PWA) [12], or browser extension. It supports many use cases including two-party call, multi-party conference, live event broadcast, and video presence. It also hides the complexity of the underlying connection, device, and media related abstractions, by exposing easy-to-use attributes or properties on the <video-io> custom element.

This paper is organized as follows. We discuss related work in Section 2. Section 3 gives motivational background, and discusses client vs. server-side app logic, and conference, call and named stream based paradigms. Section 4 describes the video-io implementation. We present interworking with the popular media servers (Janus [13], Jitsi [14], and Freeswitch [15]), and cloud hosted services (Agora [16], Firebase [17], and Livekit [18]) in Section 5. Section 6 has advanced scenarios such as video image processing, and installed PWA. Finally, our conclusions are in Section 7.

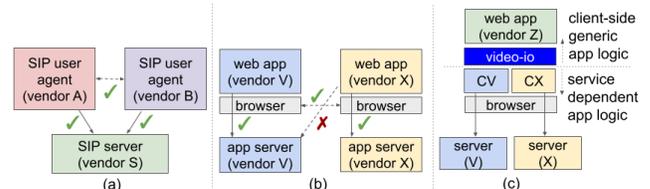

**Fig. 1** Interoperability in (a) traditional VoIP vs. (b) web-based communication systems: a web app from one vendor does not work with another. And (c) a generic web component, video-io.

## 2. RELATED WORK

Websites and web apps have used communication and media widgets embedded in web pages for a long time, such as for click-to-call, instant messaging, or audio/video players [19][20]. These widgets are apps built on the Flash Player plugin [21], or loaded in an iframe [22], or are components in a client side framework such as Angular or React, or are HTML5 standards-based web components [23]. The past attempts of such communication web components are either tied to a single vendor service [21][22][24][25], or provide only a thin feature of device capture or media playback [26][27], while using the app logic for communication on the server or elsewhere.

Our video-io web component for WebRTC is inspired by some of our earlier work on communication widgets using Flash Player or iframes [21][22], which kept the application logic in the endpoint in a resource-based software architecture [22][28], and had an extensive widget API for a wide range of use cases [21]. Flash is obsolete. Although iframes are good for independent widgets, they are cumbersome to consistently display, interact, or program with. Nevertheless, our video-io can also connect to the resource server, among other services, and hence, the plethora of example applications built using those widgets can also work with our video-io web component [29].

Previous efforts on communication interoperability have largely focused on server-side gateway, pair-wise federation among service providers, or client-side app mashups. The signaling and control of traditional communication is client driven, which made it possible to create, say, multi-protocol clients, e.g., Pidgin [30]. Although WebRTC interoperates between browsers from different vendors, it delegates its app signaling and control to the websites [9]. This allows a website owner to restrict its app users to only connect to its own service, with little or no motivation to open up the walled garden in fear of losing its users to its competitors.

Researchers have attempted to enable user reachability in such islands of communication apps by decoupling the contacts from the communication apps [7]. However, those apps are still vendor controlled, preventing independent innovation, even though many such apps have similar features of video conferencing or click-to-call. Furthermore, with the growing number [31] of such locked apps, pair-wise federation or gatewaying becomes harder.

Our goal is to promote innovation in the endpoint for video conferencing and related use cases, beyond what a single vendor locked app can offer [29][32][11]. For that, one can identify a common set of abstractions, separate from vendor specific service, so that the developers can build their web apps using those abstractions independent of a specific service. The novelty of our work is in (1) identifying and separating the app logic that runs in the browser independent of the service for many WebRTC use cases, (2) creating the first generic HTML5 web component to encapsulate that app logic, and (3) providing a pluggable connector interface to interwork with several existing and future media servers and hosted services.

## 3. BACKGROUND AND MOTIVATION

Unlike traditional static web pages, modern web apps are rich internet applications (RIA) and run complex app logic in the endpoint, while delegating only a few crucial decisions, data access, or control to the backend services. Table I lists some common app logic that runs on the client vs. server in such systems.

**Table I** Examples of endpoint vs. server driven app logic.

| | |
|---|---|
| *Endpoint driven app logic* | |
| Media capture | Capture media from microphone, camera, screen, window, file or drawing canvas. |
| Media codec | Audio/video encoder and decoder, echo cancellation, loss concealment, etc. |
| Media play | Audio/video playback, video display, scaling, padding, cropping. |
| Video layout | Multi-party video layout: grid versus presenter, talker, content share display, popout picture. |
| Bandwidth management | Tradeoff between performance and quality; adjust framerate, video size, quantization, etc. |
| Spatial audio | 3D audio placement, and stereo capture. |
| Virtual background | Blur or replace background in live video, image processing, and other video filters. |
| Annotation, whiteboard | Annotate on screen, or share emojis or reactions in video conferencing, shared whiteboard. |
| Remote control | Control of other participant's devices for remote collaboration or customer support. |
| Text chat, file, or data share | User interface for real-time text messaging, file sharing, shared notes editing, or co-browsing. |
| Encryption | End-to-end encryption of media streams, for media privacy from hosted services. |
| Reconnection | Robustness of media path connections, quick failover, and recovery. |
| Capture quality metrics | Collection of quality data or statistics, sender and receiver reports, activity logs. |
| *Server driven app logic and control* | |
| Access control | User signup, login, or access to the service. |
| Membership | Participants in a conference, with varying privileges like host, presenter or attendee. |
| Signaling | Negotiation of WebRTC signaling messages among clients, or between a client and server. |
| Media processing | Conference recording, offline transcription, notes summarization, etc. |
| Data storage | Storage of call history, shared content, quality metrics, and media recordings. |
| Relay | For the media path of clients behind restrictive firewalls or network address translators (NATs). |
| Gateway | Interwork with other systems like traditional or VoIP phone, messaging, or telepresence. |

In the context of WebRTC apps, most of the app logic including device access, connection establishment, media playback, and media stream or track management can be done in the endpoint. Servers are often needed for user reachability, signaling negotiation, centralized conference membership, access control, relay of media path for clients behind restrictive firewalls, and server-side media processing or recording. A real-world video conferencing application has a lot more app logic or features, but the general trend is the same – a vast majority of them run in the endpoint, including the crucial WebRTC.

WebRTC refers to an effort by W3C [5], IETF [6] and popular browser companies to allow web pages to exchange real-time media streams using a point-to-point media path. A web page can capture from the user's local microphone and/or camera using getUserMedia as a local media stream abstraction, create RTCPeerConnection, a peer-to-peer media path abstraction between the two browser instances, and send a MediaStream, or a collection of one or more audio/video tracks, from one browser to another, as shown in Fig. 2. Optionally, a channel abstraction, RTCDataChannel, can be used to send any data between the two browsers.

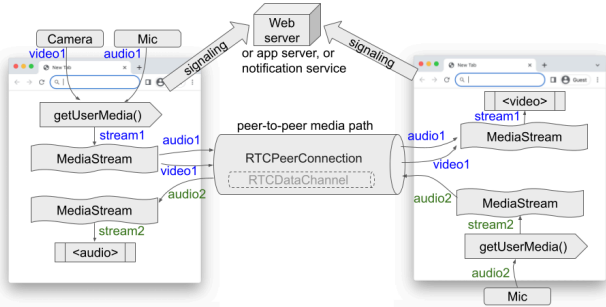

**Fig. 2** WebRTC API elements in an example session: audio/video from left to right, and audio from right to left on the same peer connection, with two media streams, one in each direction.

An RTCPeerConnection object emits certain signaling data such as session description and transport addresses, which must be sent and applied to a similar object at the other browser to establish a media path. In practice such signaling negotiation is facilitated by a vendor controlled notification service (web server, in Fig. 2), and the media path is often between a client and a vendor controlled media server, instead of directly between the clients.

Signaling and control in existing WebRTC apps are often modeled after a phone call or conference bridge abstraction [33]. In the *call* abstraction, one endpoint sends a call intent with an offered session to the other, via the notification service, and the other endpoint responds with an answered session to accept the call. In the *conference* abstraction, a conference room is identified by a URL scope or path, and all the participants connecting to the notification service under the same scope join the same conference session, and are able to communicate with each other.

A third one, called the *named stream* abstraction [33][21], was once popularized by Flash Player, but is also applicable to WebRTC apps [22][29]. In particular, it is logically similar to a NetStream object in ActionScript. Here, a client can publish and play named streams. A stream has at most one publisher and zero or more players at any time. The primary difference with the previous abstractions is that a stream is logically unidirectional. Thus, a two party call needs two streams, one published by each participant and played by the other. Similarly, a N-party conference needs N-streams. Behind the scenes, the app is often optimized using bi-directional peer connections when needed.

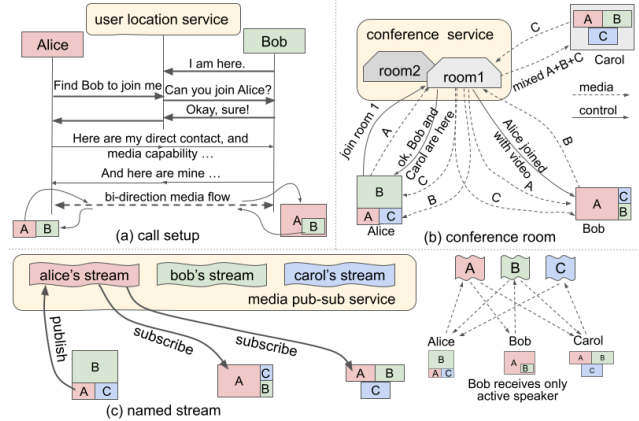

**Fig. 3** Abstractions: (a) call setup, (b) conference room, and (c) named stream. It shows an example video conference layout at each endpoint, and the control and media paths.

Fig. 3 shows the three abstractions. The call abstraction largely deals with user reachability, and only touches upon the media part, making it independent of the client-side app logic listed in Table I. The offer-answer negotiation in WebRTC appears similar to this model. The conference room abstraction can either forward or mix the participants' media. But it imposes unnecessary context and constraints, e.g., what media types are enabled by each participant, whether a participant can send multiple videos, whether participants have uni- or bi-directional media, and which server the participants connect to. Systems that have this room abstraction often mimic such constraints [41], e.g., all participants of the room must connect to the same media server, or a client must join in send-recv mode even if it has no camera, e.g., a TV app, or a user can send only one video even if she has multiple webcams, or a user cannot add another user's video box until that user publishes. This eventually leads to unnecessary rigidity in user experience, and poor reliability and scalability of the overall system.

The named stream abstraction has a proven track record in the Flash developer community. It has worked well in numerous use cases including calls and conferences. It allows all the client-side app logic listed earlier, as well as clear robustness and scalability of the server infrastructure [21][22][35]. It is also used by the popular Janus media server [13]. It is possible to create a call or conference on top of named streams (and vice versa) [33].

These reasons, and some more [29], motivated us to use the named stream abstraction in our project. We pick it as the base layer to define a clean and concise interface between the generic client side web component and the vendor specific service. Note that such a system needs a service to maintain the stream states, and to allow a client to publish or subscribe to a named stream.

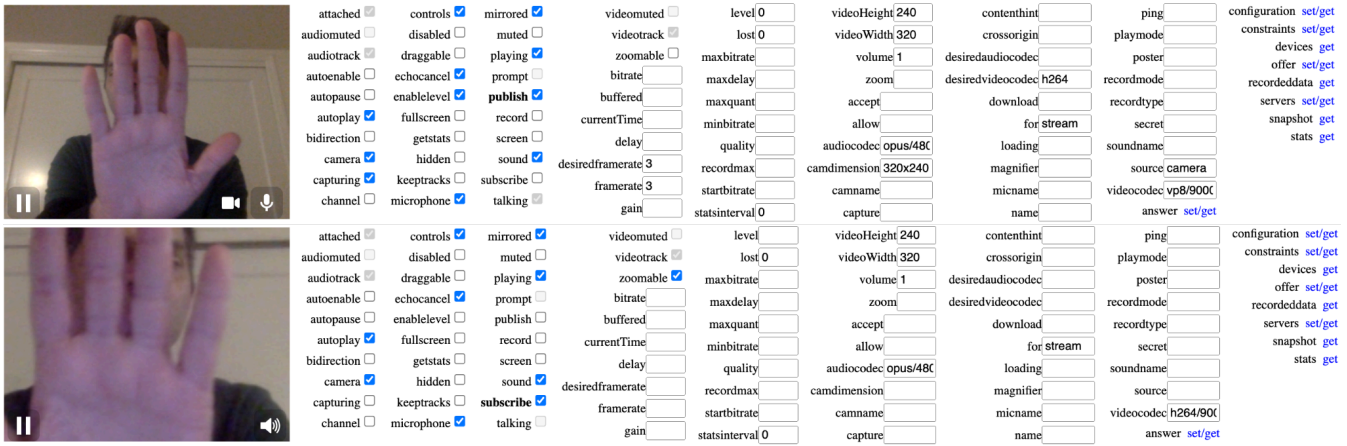

**Fig. 4** Test page with two video-io components: top one publishes, and bottom one subscribes, with a zoomed-in video. Some defaults are changed.

## 4. VIDEO-IO IMPLEMENTATION

The video-io web component can be used in many WebRTC related use cases, e.g., live camera preview, recording of multimedia messages, live video call or conference, and in client-server as well as peer-to-peer topology. It combines the various media and connection constructs available in WebRTC, and exposes a single video box abstraction. It can be embedded in a web page, say, with user controls enabled, and showing the local webcam preview as shown below.

```
<video-io controls publish="true"></video-io>
```

There are more than 80 properties, as listed in Appendix A, including the two shown above to control the behavior of the component, or to get an indication of its internal states. Many of these are also settable as attributes. Fig. 4 shows our test page that allows testing all these properties (with control based on type: boolean, number, or string), methods, and events using two connected components. Interested readers are referred to the RTC bricks project [11] for an extensive documentation of this and other related web components.

Although many of these properties are related to the media and connection state, some are used for local recording and playback, and some of them control the appearance of the component. For example, whether to show the controls for devices or play state, whether to flip horizontally the camera captured preview video, to set a poster image to display before any publish or subscribe is done, to fit the video with padding, scaling, or zooming in the component if the video aspect ratio is different than the component's display aspect ratio, to add custom buttons, text or other controls in the header or footer of the component, to show a magnifying glass, to drag-select an area for zooming into a part of the video, and so on.

For real-time media flow, a publisher instance of video-io is connected to a subscriber instance. The two instances can simply be connected in a point-to-point manner as shown below. This utilizes the WebRTC peer connection's offer-answer semantics, where one instance emits the data events, which are applied to the other. The two instances may be running on separate browsers or machines, in which case external signaling is used to exchange the data.

```
v1.addEventListener("data",e=>v2.apply(e.data));
v2.addEventListener("data",e=>v1.apply(e.data));
v1.publish = v2.subscribe = true;
```

Unlike this point-to-point connection, the named stream abstraction has many benefits, because it allows multiple subscribers on one publisher, and allows the publisher and subscribers to join and leave in any order, and any number of times. Its component design also enables integration with existing notification or media services.

Typically a video-io component is attached to a named stream component, and is configured to either publish or subscribe. Fig. 5 shows how the video-io and named stream components work together: from one publisher to one subscriber, from one publisher to many subscribers in a one-to-many broadcast, or between the endpoints in a two party call.

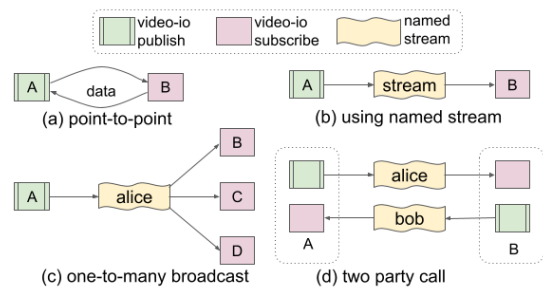

**Fig. 5** Example interactions between video-io and named stream.

The named-stream web component implements the basic necessary interface for this abstraction, as illustrated below, with one publisher and two subscriber video-io instances.

```
<named-stream id="s1"></named-stream>
<video-io for="s1" publish="true"></video-io>
<video-io for="s1" subscribe="true"></video-io>
<video-io for="s1" subscribe="true"></video-io>
```

Some of the properties, methods and events of the video-io component form the interface with the named stream component, e.g., the list of peer connections, the media stream, the underlying video element, and the signaling data. The named stream interface that is used by the video-io component includes methods to publish, subscribe, stop, and to add or remove media tracks. Besides these, it can have other service specific properties such as to locate the server, or to identify a specific stream.

Fig. 6 shows the two interfaces of video-io, one (labeled, A) is used to create web audio/video communication apps, and the other (S) is used by the named stream implementations. These are further described in Appendix A.

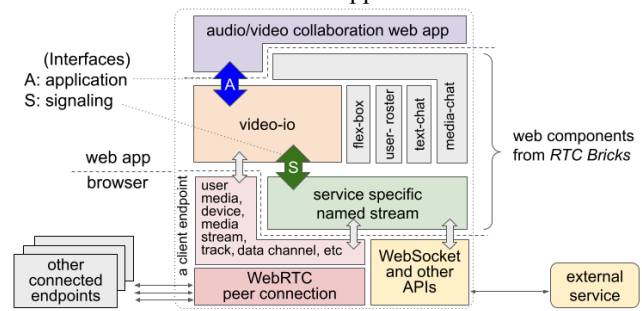

**Fig. 6** Interface among various web components in the software architecture, connecting to the external service or other endpoints.

Now, named-stream is an abstract component that enables testing and local demonstration on a single web page. In a real application, to support distributed publishers and subscribers, this is replaced by a service-specific one with the same interface. In that case, the component instance uses additional properties such as src to point to the same named stream, even if embedded in web pages of different machines. An example of firebase-stream using the Firestore real-time database as the service is shown below. Here the identifier, 1234, in the src attribute identifies the specific stream, and other config values help authenticate with the service.

```
<firebase-stream src=""></firebase-stream>
<video-io publish="true"></video-io>
<script type="text/javascript">
 document.querySelector("firebase-stream").src
  = "id:1234?config="
    + encodeURIComponent(JSON.stringify({
       apiKey: ..., projectId: ..., appId: ...
    }));
</script>
```

We discuss the implementations of such service-specific named streams in the next section.

## 5. NAMED STREAM IMPLEMENTATIONS

The primary purpose of a named stream service is to allow exchanging signaling data between a publisher and a subscriber. For each such service, we implement a new web component derived from the named stream abstraction, and connect it to that external service to provide the same well defined interface. Table II lists the various such services, and their web components implemented in our system.

**Table II** Named stream components and their external services.

| Data storage based services | |
|---|---|
| rtclite-stream | RTCLite's notification server |
| firebase-stream | Google Cloud's Firestore database |
| storage-stream | Shared storage: local or remote |
| dbgun-stream | Graph universe node database |
| *Media server or conference services* | |
| janus-stream | Open source Janus media server |
| verto-stream | Open source Freeswitch server |
| agora-stream | Agora.io RTE service |
| jitsi-stream | Jitsi-as-a-service jaas.8x8.vc |
| livekit-stream | Livekit.io RTC service |
| videosdklive-stream | videosdk.live RTC service |
| *Generic stream behavior/implementation* | |
| split-stream | Fork published stream to multiple |
| external-stream | Custom service connector in iframe |

*Data storage based services*

The RTCLite project [34][36] has a lightweight WebRTC notification service written in less than 200 lines of Python code. It provides a simple API for named stream publish and subscribe. It can also allow a publisher to hide the raw stream name, by using hashed stream names at the subscribers. The rtclite-stream web component takes a websocket URL of the named stream on the server to publish or subscribe to.

Google's Cloud Firestore [17] is a real-time database. The firestore-stream web component takes a stream name and Firebase credentials to connect to the service. It stores the publisher and subscriber states, and facilitates messages among them, using the Firestore API.

The storage-stream web component uses another web component, shared-storage [11], to store the publisher and subscriber states and to exchange messages among them. The shared-storage component is a generic real-time database abstraction that stores hierarchical structured data or resources, and provides change notification events. We have three implementations: first one uses the localStorage of the browser for testing on a single machine, the second one uses an external resource server [22][28], and the third uses a peer-to-peer network for decentralized data storage [37]. We have light weight implementations of the resource server in PHP and Python [28][38], and an ongoing effort using NodeJS.

The Graph Universal Node database [39] is an open source decentralized database. The dbgun-stream web component takes a stream name, and bootstrap node URL, and other config. It connects to and participates in the peer-to-peer data storage.

These first four web components have one thing in common. They all keep the peer connection logic inside the video-io component, not in the service, and only deal with the exchange of signaling messages. In particular, they do not handle media streams. They all create full mesh media path topology among the video-io instances on different machines, with a separate peer connection plus media stream for each publisher-subscriber pair.

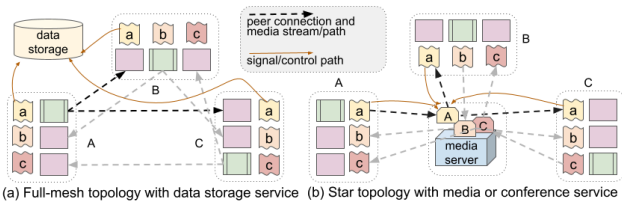

**Fig. 7** Media path topology with (a) data storage services, versus (b) media or conference services, with a room assigned to each named stream. The signaling/control path is centralized in both; and control path only of only one stream, A, is shown for brevity.

Unlike these, the next six web components we describe keep the peer connection logic as part of the specific named stream web component implementation. Some of the connection dependent properties in the application interface (A in Fig. 6) of video-io are not supported if using a vendor specific stream implementation. These also have a centralized media server, resulting in a star media path topology. The difference is shown in Fig. 7, where the data storage services include the first four, and the media or conference services are the next six as shown in Table II.

*Media servers or conference services*
The underlying WebRTC technology used in the video-io component is inherently peer-to-peer for the media path. However, many real-world call and conferencing applications rely on media servers for various functions including recording, interactive voice prompts, audio mixing, or video switching or routing.

Janus [13] is a popular open source and general purpose WebRTC media server. It supports the publisher-subscriber connection in its videoroom plugin, and implements video forwarding, or selective forwarding unit (SFU). Our janus-stream web component takes a websocket or http URL containing the server location, and room and feed identifiers as parameters, to identify a specific stream in a specific room, on the server.

Freeswitch [15] is a popular open source telephony server that supports voice, video, text using traditional VoIP such as SIP [8] as well as newer WebRTC. It implements server side mixing, or multipoint control unit (MCU) for audio and video. Our verto-stream web component takes a websocket URL to connect to the server, and uses its API over a remote procedure call (RPC) interface to login, and create a session in a conference. A named stream maps to a unique session at the server, and all connected video-io instances join the same session. The client side logic works together with the server side processing of a connection to that session, such that the session identifier is used to create a separate conference if needed.

Jitsi [14] is another open source video conferencing server, which also has a hosted jitsi-as-a-service offering. Our jitsi-stream web component takes a room identifier representing the stream name, an application identifier and an authentication token, to connect to this service.

Similarly, Agora [16], LiveKit [18], and Video SDK Live (videosdk.live) are developer friendly hosted video call and conferencing services that expose a conference room abstraction. The agora-stream component takes a channel name representing a named stream, and an application identifier and authentication token. The livekit-stream component takes a room name representing a named stream, and a sandbox test server to connect to. The videosdklive-stream component takes a room name for a named stream, and an authentication token that includes other data for connection to that service.

Even though these services all implement a conference room abstraction, there are significant differences in the APIs on how to connect, authenticate, and send or receive a media stream. These differences make a strong case for a project like ours that attempts to isolate such differences from the core client side application logic. Our specific named stream implementations hide the differences, and expose a simple and uniform API, to the generic video-io component. The endpoint applications built using video-io can readily switch to a different media server or hosted service, by replacing the named stream implementation. Web developers or hosted service vendors can create such named stream implementations as web components for their developer community.

*External stream component*
Additionally, we provide a generic external-stream web component, that takes a URL of an external web app, loads it in an internal sandboxed iframe or an external popup window, and delegates the named stream interface function to that web app as shown in Fig. 8. Since Chrome does not support transfer of a MediaStream object between window contexts, we use a peer connection between the main web application and the external web app, to exchange the media streams - the published media stream from the video-io is sent from the main app (at the top, in Fig. 8) to the external web app in the iframe or window (at the bottom), and the subscribed media stream is sent from the external web app to the main app.

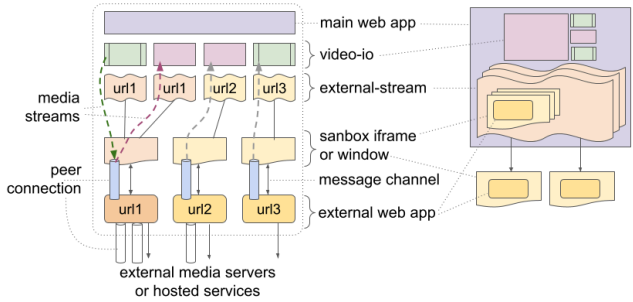

**Fig. 8** Interaction among various elements when using the external-stream web component. A stream name/url maps to an external web app loaded in a sandbox iframe or external window.

The controlled cross origin communication between the loaded iframe or window, and the main window ensures that the main web app is protected from any malicious or unintentional behavior of the external web app loaded in the sandboxed iframe or external window. This approach allows us to keep such service-specific named stream implementations outside our code, and potentially hosted by the same vendor that provides the service.

*Split stream and video-io fork*

The named stream abstraction allows at most one publisher and zero or more subscribers attached to a single stream. It also requires a video-io component instance to be attached to only one named stream. Note that (see Appendix A) the video-io component has a reference to a named stream via the for or srcObject property, and the named stream's publish and subscribe methods get the video-io's reference as a parameter. Except for the base named-stream component, all the other derived named stream components (shown in Table II), are usually attached to by only one video-io instance. This is because a real video conference app endpoint needs only one video-io instance for each named stream, and other video-io instances for that same named stream are on other endpoints.

A subscriber video-io attached to one named stream is fine, but a publisher video-io attached to only one named stream can be limiting for some application use cases, where a single video-io may need to be published to multiple streams. There are two ways to do this: use video-io to fork, or use split-stream, as shown in Fig. 9.

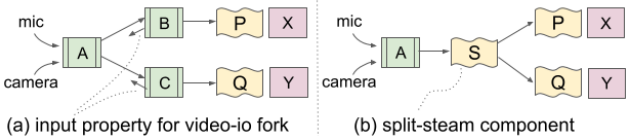

**Fig. 9** How to publish one video-io to multiple named streams?

In the first case, two new video-io components are used, with their input properties set as the first video-io publisher. In the second case, the split-stream web component contains two or more other named stream components, and allows the split-stream to use the published media stream in both the included named streams. The publisher video-io attaches to the parent split-stream component, whereas the subscriber video-io attaches to the individual child named stream.

The video-io fork and the split-stream web component can also allow creating advanced media path topologies in the web app, such as for three party calls, or application level multicast.

## 6. ADVANCED USE CASES

We describe some advanced applications of the video-io web component, e.g., to modify a video in realtime, add caption, or perform image processing on each frame. These can be done in either publish or subscribe mode. In the subscribe mode, it is also possible to republish a clone of the received video with or without modification.

The input property mentioned earlier can be set as either a video or canvas element, or a MediaStream instance, or another video-io or a video-mix instance. When the input property of a publisher video-io component is set as a video element, it uses that video element as the media source, instead of capturing from a camera or microphone. This is used for streaming a video file to a named stream, e.g., to play an external or uploaded video in a conference.

The video-io components and the named streams can be cascaded to build a peer-to-peer broadcast topology without using a media server, as shown in Fig. 10.

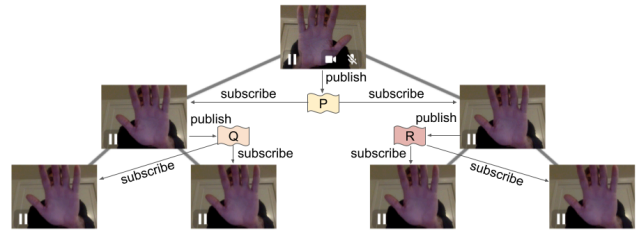

**Fig. 10** Peer-to-peer media broadcast. Pausing an intermediate video pauses all the videos in the subtree rooted at that node.

A video-mix web component combines videos from one or more sources, by using a canvas element to create a mixed video layout, or a modified video stream. It periodically calls the application supplied script to modify and render an image frame on the canvas. This component can be used to manipulate the captured webcam stream in real time such as to add caption using speech-to-text of microphone audio, add overlay logo or text watermark, perform image processing, detect and blur background, or replace background with an image or another video. It can also be used to combine multiple videos, e.g., for a video layout in a multiparty call, or to include webcam preview as picture-in-picture in the shared screen video feed. Fig. 11 illustrates some of these client side app logic. Many endpoint driven innovations enabled by the RTC Helper [32] browser extension are also possible with video-io, directly in the main web app without a browser extension.

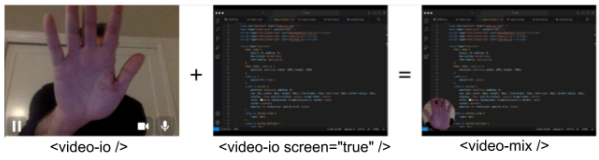

(a) combined video is webcam video overlaid on screen capture video

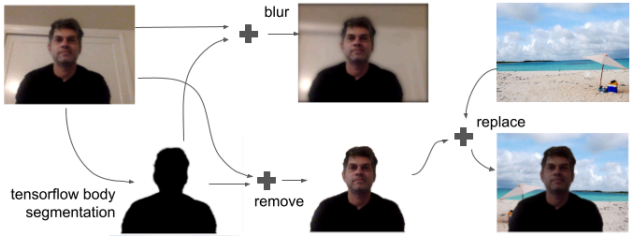

(b) video background detection, removal, blur, or replace.

**Fig. 11** Using video-mix to combine and/or modify videos.

Some of the video-io properties are used to record or play video messages or files. The recording is done in the client side browser. It also supports continuous recording, say, of the last 30 seconds, to allow a conference participant to quickly play back the last few seconds in a call. Additionally, its delayed playback feature helps in real-time captioning, and live event broadcasts with moderation.

The secret property of the video-io component is used to quickly enable end-to-end encryption using WebRTC's insertable streams feature [40]. This works when the media servers or hosted services do not need access to the decoded audio/video data.

Although video-io is a web component, intended to show a single video in publish or subscribe mode, many video conferencing related use cases require showing multiple video elements at the same time. The flex-box web component is used to layout multiple video-io or other elements. It has properties to control the layout, e.g., grid vs. presenter mode, or pagination control.

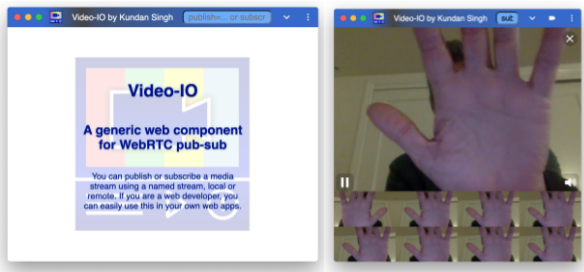

**Fig. 12** Screenshots of video-io PWA: empty container on the left; and video-io instances attached to the same named stream on right.

We have also created a standalone web app named video-io that can show zero or more video-io web components wrapped in a flex-box component, as shown in Fig. 12. It can be installed as a PWA [12], on desktop or mobile. The user interface allows publish or subscribe of named streams. It attaches a video-io component to each such stream. It can also be launched on desktop, if not already open, to add a new stream using its URL protocol handlers: web+ezpub or web+ezsub, for publish or subscribe, respectively. If a user opens a URL web+ezpub:rtclite:wss://example.com/str/15, it adds a video-io instance, and an attached rtclite-stream web component. It then connects to the presumed rtclite service at example.com over a secure websocket transport, and publishes camera/mic media to the named stream, str/15.

The flex-box component allows dynamically adjusting the layout, e.g., double clicking on a video to put it in spotlight, or resize videos based on available size and aspect ratio of the window. It allows drag-and-drop of the video elements, e.g., to re-order them in the display, or to pop out a video to a separate window or tab, or to move or copy a publish or subscribe video stream from one app instance to another. These usability features are similar to that of Vclick [35], but are not commonly found in popular video conference apps.

The goal of this video-io app, separate from the video-io web component, is to create a generic user interface for web video conferencing applications, where the conferencing app logic can reside in external software, which invokes this app to display, capture, publish and subscribe various media streams using the protocol handler. The extensive video-io component API can be used to further customize its behavior using the URL parameters.

## 7. CONCLUSIONS AND FUTURE WORK

We have presented video-io as a generic web component that separates the client side application logic from the vendor specific services. It enables innovation in the web app running in the endpoint for a wide range of audio/video communication use cases, independent of the vendor service restrictions or constraints. Our implementations of the named stream abstraction not only caters to the existing popular media servers and hosted services, but also allows external custom implementations for other services.

With the growing number of WebRTC enabled video call and conferencing services, we continue to see the trend of vendor lock-ins. The end result is that only a few popular apps dictate the experience of a vast majority of the users. Our attempt is to separate the service specific part from the client app logic. It allows web developers to create such communication apps independent of any specific service, and to easily connect it to a wide range of media servers or hosted services. Our video conferencing app allows participants on different vendor services to join the same call, and does application mashups in the endpoint.

Our work promotes reuse of media features across web apps, reduces vendor lock-in of communication features, and provides a flexible, extensible, secure, feature-rich, and end-point driven web component to support a large number of use cases for audio/video apps. One could argue that WebRTC is a step closer to avoiding vendor lock-in. We argue that it is not close enough, and in fact worsens the situation due to its lack of signaling or control in the specification. One could argue that vendor lock-in is as expected or intended, and what is good for an individual business is good for the technology space in general. We argue against this, because we believe that independent development of endpoints or apps from communication or network services promotes more innovation.

We plan to work on cross browser compatibility, expansion of our web components inventory, optimization and reuse of peer connections for multiple streams, support for persistent and/or non-media content in a named stream, and applying video-io in more novel use cases promoting innovation in the endpoint. We also plan to extend our installed app to support gesture and voice control as user input, and to allow external call signaling using push notification or SIP [8], or from third-party web apps using browser extensions [32].

# APPENDIX A: VIDEO-IO REFERENCE

The complete reference for the properties and attributes of the `video-io` web component in the application interface (labeled A in Fig. 6) is included below. A few of the properties are not yet implemented, and marked as such. All the `boolean`, `string` and `number` properties can be used as attributes too. But a `boolean` property that has an `attribute` marking below does not need a value, and the presence of the attribute indicates that the property is true. This is to make them similar to existing attributes in `video` elements.

Table III  Properties of video-io application interface.

| Name | Type and description |
|---|---|
| attached | bool (real-only), default false, indicate |
| | Whether the component instance is attached to DOM? |
| audiomuted | bool (real-only), default false, indicate |
| | Whether the audio track is muted on subscribe? |
| audiotrack | bool (real-only), default false, indicate |
| | Whether an audio track exists on subscribe? |
| autoenable | bool, default false, control |
| | Auto-enable/disable outbound track on play or pause. |
| autopause | bool, default false, control |
| | Publisher informs subscriber to pause/play before actual pause/play to avoid black video on subscriber. |
| autoplay | bool, default true, control |
| | Underlying video's autoplay, on publish. |
| camera | bool, default true, control |
| | Use a webcam video capture device and outbound track. |
| capturing | bool, default false, control+indicate |
| | Whether outbound tracks are enabled on publish. |
| channel | bool default false, control |
| | Establish a data channel after publish or subscribe. |
| controls | bool, default false, attribute, control |
| | Display user interface controls for publish or play. |
| disabled | bool, default false, attribute, control |
| | Disable the user interface controls. |
| draggable | bool, default false, attribute, control |
| | Allow drag and drop, with data as properties in text. |
| echocancel | bool, default true, control |
| | Enable echo cancellation in microphone capture. |
| enablelevel | bool, default false, control |
| | Periodically update `talking` and `level` to indicate sound level in microphone on publish or received audio on play. |
| fullscreen | bool, default false, control+indicate |
| | Whether the component is in full screen display? |
| getstats | bool (write-only), default false, control |
| | Calls getStats on all peer connections, and updates bitrate, quality, audiocodec, etc., immediately. |
| hidden | bool, default false, attribute, control |
| | Hide the display, useful for audio-only use cases. |
| keeptracks | bool, default false, control |
| | Do not remove or add a track when the `camera` or `microphone` is updated, but just disable or enable the outbound track. |
| microphone | bool, default true, control |
| | Use an audio capture device and outbound audio track. |
| mirrored | bool, default true, control |
| | Whether to flip horizontally the video on publish? |
| muted | bool, default false, attribute, control+indi |
| | Whether the microphone or sound is in mute state? |
| playing | bool, default false, control+indicate |
| | Whether the media is playing in publish or subscribe? |
| prompt | bool (read-only), default false, indicate |
| | Whether the device capture permissions prompt is shown? |
| publish | bool, default false, control |
| | Set this component in publish mode. |
| record | bool, default false, control+indicate |
| | Whether this is getting recorded in publish or subscribe? |
| screen | bool, default false, control |
| | Use screen capture instead of webcam in publish? |
| sound | bool, default true, control+indicate |
| | Whether to use an audio play device on subscribe? |
| subscribe | bool, default false, control |
| | Set this component in subscribe mode. |
| talking | bool (read-only), default false, indicate |
| | Whether active talking is detected in publish or subscribe? |
| videomuted | bool (read-only), default false, indicate |
| | Whether the video track is muted on subscribe? |
| videotrack | bool (read-only), default false, indicate |
| | Whether a video track exists on subscribe? |
| zoomable | bool, default false, control |
| | Allow zooming in using drag-select, or click to revert. |
| bitrate | number (read-only), indicate |
| | The current total bitrate in kb/s. |
| buffered | number (read-only), indicate, not impl. |
| | The current buffer size in seconds. |
| currentTime | number (read-only), indicate, not impl. |
| | The current play time in seconds. |
| delay | number (read-only), indicate, not impl. |
| | The current media delay in seconds. |
| desiredframerate | number, control |
| | Desired video framerate in frames-per-second in publish. |
| framerate | number (read-only), indicate |
| | The current framerate in frames-per-second. |
| gain | number, default 1, control+indicate |
| | The sound level of microphone capture, between 0.1 to 10 (multiplier), corresponding to -20dB to +20dB (linear). |
| level | number (read-only), indicate |
| | The current audio/sound level. |
| lost | number (read-only), indicate |
| | The cumulative packet lost count. |
| maxbitrate | number, control |
| | The maximum bitrate in kb/s for video in publish. |
| maxdelay | number, control, not impl. |
| | The maximum delay allowed in subscribe. |
| maxquant | number, control |
| | The maximum video encoder quantization parameter in publish, low value is better quality, applies only to VP8. |
| minbitrate | number, control |
| | The minimum bitrate in kb/s for video in publish. |
| quality | number (read-only), indicate, not impl. |
| | Quality score, from 0 (worst) to 1 (best). |
| recordmax | number, control |
| | Maximum duration in seconds of the recording. A negative number keeps the last N-seconds of recording. |
| startbitrate | number, control |
| | The starting bitrate in kb/s of video in publish. |
| statsinterval | number, control |
| | Interval in seconds to periodically call `getStats` internally to update various other properties. |
| videoHeight | number (read-only), indicate |
| | Height in pixels of the video currently showing. |
| videoWidth | number (read-only), indicate |
| | Width in pixels of the video currently showing. |
| volume | number, control+indicate |
| | The sound level of the audio play, of the underlying video. |
| zoom | number, default 1, control |
| | Multiplier to zoom or scale the controls, header, footer. |
| answer | object, control |
| | Constraints for creating answer SDP. |
| configuration | object (write-only), control |
| | Optional configuration to create a peer connection. |
| constraints | object (write-only), control |
| | Optional constraints to create a peer connection. |

 

| Property | Type | Description |
|---|---|---|
| devices | object (read-only) | Returns a promise that resolves to a list of devices, using the underlying enumerateDevices API. |
| input | object, control | Set this to an external video, canvas, another video-io or video-mix, as source to publish, instead of using cam/mic. |
| offer | object, control | Constraints for creating offer SDP. |
| recordeddata | object (read-only) | The blob or a promise of a blob of recorded data so far. Reset the data after access, unless recordmode is "append". |
| servers | object, control | Array of ICE servers to use for peer connection. |
| snapshot | object (read-only) | Snapshot image as a data URL string, captured from the displayed video at the time this is accessed. |
| srcObject | object, control | The named stream component to use. See "for". |
| stats | object (read-only) | Returns a promise that resolves to the result of getStats on all the peer connections, as an array or arrays. |
| video | object (read-only) | A reference to the underlying video element. |
| accept | string, attribute, control, not impl. | The content type that can be played or recorded. |
| allow | string, attribute, control, not impl. | A list of features that are allowed in this component. |
| audiocodec | string (read-only), indicate | The audio codec in use, e.g., "opus/48000". |
| camdimension | string, control | Set the desired camera capture size, e.g., "640x360". |
| camname | string, control | The desired camera device identifier to publish. |
| capture | string, control, not impl. | The type of capture device to use. |
| contenthint | string, control | The same property of the underlying track in publish. |
| crossorigin | string, attribute, control, not impl. | The cross origin resource sharing behavior. |
| desiredaudiocodec | string, control | The audio codec for publish or subscribe, with wildcard. |
| desiredvideocodec | string, control | The video codec for publish or subscribe, with wildcard. |
| for | string, attribute, control | The id of a named stream component attached to this. |
| magnifier | string, control | Whether to show a magnifier on mouse hover, controlling the size and scale. |
| micname | string, control | The desired microphone device identifier to publish. |
| name | string, attribute | Same as the standard name attribute. |
| ping | string, control | The webhook URLs, space separated, that receive the events, e.g., publish, subscribe, or stop. |
| playmode | string, control, not impl. | The play mode in subscribe, e.g., live or web. |
| poster | string, control | The picture to be displayed before publish or subscribe is done. |
| recordmode | string, control | Whether to overwrite (default), or "append". |
| recordtype | string, default "video/webm" control | The content type of the recorded data. |
| secret | string, control | Enable end-to-end encryption using insertable streams, using this value as the encryption key. |
| soundname | string, control | The desired speaker or audio output device identifier. |
| source | string (read-only), indicate | The media source is one of "input", "screen", "camera" or unassigned yet, in publish mode. |
| videocodec | string (read-only), indicate | The video codec in use, e.g., "vp8/90000". |

A list of events dispatched by the component is shown below. Each event object has a type attribute, and any additional attributes as mentioned below.

**Table IV** Events dispatched by video-io.

| Event | Attributes and description |
|---|---|
| propertyChange | property, oldValue, newValue |
| | When any property changes, except for write-only. |
| data | data, pc |
| | It generates new signaling data on a peer connection (pc), that should be sent out to the other end. |
| play or pause | |
| | When the underlying video element dispatches this event. |
| message | data |
| | When some data is received on the data channel. |

A list of functions and properties of video-io that are used by the named stream implementation on the signaling interface (labeled S in Fig.6) is shown below.

**Table V** Functions/properties of video-io used by named stream.

| Function/property | Signature and description |
|---|---|
| metadata | object |
| | Contains list of properties, events, methods. |
| apply | apply(...) |
| | Supply the data from the data event on the other end. |
| createPeerConnection | createPeerConnection(...) |
| | Create a peer connection based on the component state. |
| pc | Array |
| | An array of peer connection objects stored here. |
| localStream | MediaStream or null (read-only) |
| | Local capture stream in publish. |
| remoteStream | MediaStream or null (write-only) |
| | Remote received stream in subscribe. |
| send | send("...") |
| | Create a data channel, if not already, and send the text. In the case of a named stream, it can be sent from the publisher to all subscribers, or from the subscriber to the publisher. |

A list of functions and properties of the named stream, also part of the same signaling interface (labeled S), is shown below, and must be implemented by other vendor specific named stream implementations to work with video-io.

**Table VI** Functions of the named stream used by video-io.

| Function | Signature and description |
|---|---|
| publish | publish(...video-io instance) |
| | Called to publish this stream. |
| subscribe | subscribe(...video-io instance) |
| | Called to subscribe to this stream. |
| stop | stop(...video-io instance) |
| | Called to stop a previous publish or subscribe. |
| addTracks | (optional) addTracks(... array of tracks) |
| | Add new tracks to the published stream. On the other end, it should update the remoteStream property. |
| removeTracks | (optional) removeTracks(... array of tracks) |
| | Remove existing tracks from the published stream. On the other end, it should update the remoteStream property. |